\documentclass[prc,aps,groupedaddress,preprintnumbers,twocolumn]{revtex4}
\usepackage{graphicx} 

\renewcommand{\vec}[1]{{\mathbf{#1}}}

\begin{document}

\title{$0\nu\beta\beta$ and $2\nu\beta\beta$ nuclear matrix elements, QRPA, and isospin symmetry restoration}

\author{Fedor \v{S}imkovic}
\email[]{fedor.simkovic@fmph.uniba.sk}
\affiliation
        {\it  BLTP, JINR, 141980 Dubna, Moscow region, Russia
and Comenius University, Mlynsk\'a dolina F1, SK--842 48
Bratislava, Slovakia and IEAP CTU, 128--00 Prague, Czech Republic}
\author{Vadim Rodin \footnote{Now at Stuttgart Technology Center, Sony-Deutschland GmbH, D-70327, Stuttgart, Germany}}
\email[]{vadim.rodin@uni-tuebingen.de}
\affiliation{Institute f\"{u}r Theoretische Physik der Universit\"{a}t
T\"{u}bingen, D-72076 T\"{u}bingen, Germany}
\author{Amand Faessler}
\email[]{Amand Faessler <faessler@uni-tuebingen.de>}
\affiliation{Institute f\"{u}r Theoretische Physik der Universit\"{a}t
T\"{u}bingen, D-72076 T\"{u}bingen, Germany}
\author{Petr Vogel}
\email[]{pvogel@caltech.edu}
\affiliation
        {\it Kellogg Radiation Laboratory and Physics Department, 
Caltech, Pasadena, CA 91125, USA}

\date{\today}

\begin{abstract}
Within QRPA we achieve partial restoration of the isospin
symmetry and hence fulfillment of the requirement that the
$2\nu\beta\beta$ Fermi matrix element $M^{2\nu}_F$ vanishes,
as it should,
unlike in the previous version of the method. This is accomplished
by separating the renormalization parameter $g_{pp}$ of the
particle-particle proton-neutron interaction into the isovector
and isoscalar parts. The isovector parameter $g_{pp}^{T=1}$ need
to be chosen to be
essentially equal to the pairing constant $g_{pair}$, so no new
parameter is needed. For the $0\nu\beta\beta$ decay the Fermi
matrix element $M^{0\nu}_F$ is substantially reduced, while the
full matrix element $M^{0\nu}$ is reduced by $\approx$ 10\%. We
argue that this more consistent approach should be used from now
on in the proton-neutron QRPA and in analogous methods.
\end{abstract}

\maketitle

\section{Introduction}\label{S:intro} 

Answering the  questions whether  total lepton number is a conserved quantity
or not, and thus whether neutrinos
are massive Majorana fermions, is a crucial part of the search for the
``Physics Beyond the Standard Model''. Consequently, experimental searches for the 
$0\nu\beta\beta$
decay are pursued worldwide (for a recent
review of the field, see e.g. \cite{AEE}). However, interpreting existing results 
and planning new experiments
is impossible without the knowledge of the corresponding nuclear matrix elements.

The nuclear matrix elements $M^{0\nu}$ of the  $0\nu\beta\beta$ decay must be 
determined using nuclear structure theory, and the choice of the appropriate approximations 
is a crucial part of that task. Some of the methods employed for evaluation of the
 $M^{0\nu}$, in particular those that begin with the transformation from particles
 to quasiparticles to account for the like-nucleon pairing 
 ( see e.g. \cite{us1,us2,us3,us5, GCM, PHFB,IBM2}), use
 wave functions that do not exactly conserve the particle number. The number
 of protons and neutrons is usually conserved on average or, in some
 cases, it is restored by the particle number projection. In either case, until now no attempt
 was made to check that the isospin, which is known to be,
 to a very good approximation,  valid quantum number in nuclei, remains as such
 in the resulting wave functions that are obtained by solving the corresponding equations
 of motion.
 
 It is well known that by the proper treatment of the quasiparticle interaction the broken
 symmetries can be restored. Naturally, exact calculation would restore the broken
 symmetries exactly. However, even with the approximate, RPA-like treatment,
 it is possible to partially restore some of the broken symmetries. In this work
 we show, following basically the suggestions made initially in Ref.\cite{Vadim11},
 how this can be done in the case of isospin and by doing that
 the values of the Fermi nuclear matrix elements, both
 for the $2\nu\beta\beta$ and   $0\nu\beta\beta$ decays, are 
 substantially modified. Even though
 the resulting total  $M^{0\nu}$ nuclear matrix elements are changed only by
 $\approx$ 10\%, it is worthwhile, and certainly more consistent, to use in future
 the prescriptions described below.

\section{Formalism}\label{S:form}

Assuming that the  $0\nu\beta\beta$ decay is caused by the exchange of light Majorana neutrinos, the
half-life and the nuclear matrix element are related through
\begin{equation}
\frac{1}{T_{1/2}^{0\nu}} = G^{0\nu}(Q,Z) | M^{0\nu} |^2 ~|\langle m_{\beta\beta} \rangle |^2 ~,
\end{equation}
where $ G^{0\nu}(Q,Z)$ is the calculable phase space factor, $\langle m_{\beta\beta} \rangle$ is
the effective neutrino Majorana mass whose determination is the ultimate goal of the experiments,
and $ M^{0\nu}$ is the nuclear 
matrix element consisting of Gamow-Teller, Fermi and Tensor parts,
\begin{equation}
 M^{0\nu} = M^{0\nu}_{GT} - \frac{M^{0\nu}_F}{g_A^2} +  M^{0\nu}_T \equiv   
M^{0\nu}_{GT} ( 1 - \chi_F/g_A^2 + \chi_T ) ~,
\label{eq:mgt}
\end{equation}
where $ \chi_F$ and $ \chi_T$ are the matrix element ratios
$\chi_F = M^{0\nu}_F/M^{0\nu}_{GT}$ and 
$\chi_T = M^{0\nu}_T/M^{0\nu}_{GT}$. (In the literature a different notation
is sometimes used, $\chi_F = M^{0\nu}_F/(g_A^2 M^{0\nu}_{GT}$.)

 The main GT part, $ M^{0\nu}_{GT} $,  can be somewhat symbolically written as
\begin{equation}
 M^{0\nu}_{GT} = \langle f | \Sigma_{lk} \vec{\sigma}_l \cdot  \vec{\sigma}_k \tau_l^+ \tau_k^+ 
H(r_{lk},\bar{E}) | i \rangle ~,
\end{equation}
where $ H(r_{lk},\bar{E})$ is the neutrino potential described in detail in \cite{us5} and $r_{lk}$ is the relative
distance between the two neutrons that are transformed in the decay into the two protons.

Analogously, the Fermi matrix element is
\begin{equation}
 M^{0\nu}_F = \langle f | \Sigma_{lk} \tau_l^+ \tau_k^+  H(r_{lk},\bar{E}) | i \rangle ~.
\end{equation}
Note that these $0\nu\beta\beta$ matrix elements are expressed in the closure approximation; its applicability
is also discussed in \cite{us5}. However, the results reported later in this work were obtained without
using the closure; instead explicit summation over all virtual intermediate states was performed.

The half-life of the experimentally well studied $2\nu\beta\beta$ decay depends formally on two nuclear matrix elements 
\begin{equation}
\frac{1}{T_{1/2}^{2\nu}} = G^{2\nu}(Q,Z) [ M^{2\nu}_{GT} + \frac{g_V^2}{g_A^2}M^{2\nu}_F]^2.
\end{equation}
The Gamow-Teller $2\nu\beta\beta$ matrix element is
\begin{equation}
M^{2\nu}_{GT} = \Sigma_m \frac{ \langle f || \Sigma_k \sigma_k \tau^+_k || m \rangle \langle m || \Sigma_l \sigma_l \tau^+_l || i \rangle}
{E_m - (M_i + M_f)/2} ~,
\label{eq:2nu}
\end{equation}
where the summation extends over all $1^+$ virtual intermediate states. 
In that case the closure approximation is not a valid approach but can be formally introduced  
by defining  the corresponding closure matrix element $M^{2\nu}_{cl}$ when replacing the energies 
$E_m$ by the proper average value $\bar{E}_{2\nu}$.
Thus,
\begin{eqnarray}
M^{2\nu}_{GT(cl)} \equiv  \langle f | \Sigma_{lk} \vec{\sigma}_l \cdot  \vec{\sigma}_k 
\tau_l^+ \tau_k^+ | i \rangle~,
\nonumber \\
M^{2\nu}_{GT(cl)} = M^{2\nu} \times (\bar{E}_{2\nu-GT} - (M_i + M_f)/2) ~.
\label{eq:2nucl}
\end{eqnarray}

Formally, in the description of the $2\nu\beta\beta$ decay also appears the Fermi matrix element
\begin{equation}
M^{2\nu}_F = \Sigma_m \frac{ \langle f ||  \Sigma_k \tau^+_k || m \rangle \langle m || \Sigma_l \tau^+_l || i \rangle}
{E_m - (M_i + M_f)/2} ~,
\label{eq:2nuF}
\end{equation}
where the summation extends over all $0^+$ virtual intermediate states, and its closure form is
\begin{eqnarray}
M^{2\nu}_{F(cl)} \equiv  \langle f | \Sigma_{lk} \tau_l^+ \tau_k^+ | i \rangle~,
\nonumber \\
M^{2\nu}_{F(cl)} = M^{2\nu}_F \times (\bar{E}_{2\nu-F} - (M_i + M_f)/2) ~.
\label{eq:2nuFcl}
\end{eqnarray}
The ground state $| i \rangle$ of the initial nucleus has isospin $T \equiv T_z = (N-Z)/2$ while the final
state $| f \rangle$ has isospin $T-2 \equiv T_z = (N-Z - 4)/2$.
Since the operator $ \Sigma_k \tau^+_k$ just changes the isospin projection and cannot change
the total isospin, it is obvious that when isospin is a good quantum number both
Fermi matrix elements must vanish,
\begin{equation}
M^{2\nu}_F = M^{2\nu}_{F(cl)} = 0,
\label{eq:2nuF0}
\end{equation}
since the average energy denominators in Eq. (\ref{eq:2nuFcl}) are nonvanishing.

Until now, within QRPA, PHFB, EDF and IBM-2 methods the validity of condition Eq.(\ref{eq:2nuF0}) has
not been usually tested (\cite{us1,us2,us3,us5, GCM, PHFB,IBM2}). 
When results were published, $M^{2\nu}_F$ and $M^{2\nu}_{F(cl)}$ do not vanish
and are, in fact, comparable to $M^{2\nu}_{GT}$ and $M^{2\nu}_{GT(cl)}$,
respectively. Despite that, when evaluating the $2\nu\beta\beta$ half-life the
Fermi matrix element was usually simply neglected. 

As we show further, in the usual application
of QRPA the condition  Eq.(\ref{eq:2nuF0}) is not obeyed. Instead, the magnitude of $M^{2\nu}_F$
is numerically comparable to  the magnitude of $M^{2\nu}_{GT}$ as just pointed out. 
In addition, for the $0\nu\beta\beta$ decay, within QRPA the ratio $\chi_F \approx -0.5$
while in the nuclear shell model, where isospin is a good quantum number by construction, the condition
Eq.(\ref{eq:2nuF0}) is, naturally, obeyed and $\chi_F \approx -(0.2-0.3)$ \cite{ISM}.

Where does this problem in QRPA method originate?
The method begins with the Bogoliubov transformation relating the
particle creation and annihilation operators $a^{\dagger}_{jm}, \tilde{a}_{jm}$
with the quasiparticle creation and annihilation operators  $c^{\dagger}_{jm}, \tilde{c}_{jm}$.
By solving the BCS equations one includes the neutron-neutron and proton-proton 
isovector pairing interactions.

At this stage several symmetries are broken. The numbers of protons $Z$ and neutrons $N$ 
are no longer exact, but valid only on average. In addition, since the neutron-proton part
of the isovector pairing interaction is neglected, additional source of isospin violation 
is introduced. It turns out that it is relatively easy to remedy this additional effect and
restore the isospin conservation, at least in part, as explained further here.
As the RPA ( and QRPA) is derived from the equation of motion for bifermionic operators 
(treated in the quasiboson approximation), symmetries of the model Hamiltonian can naturally 
be fulfilled in that approximation.

To proceed further, the equations of motion need to be solved.
Within the QRPA method the forward- and backward-going amplitudes $X$ and $Y$ that are needed for the evaluation
of the nuclear matrix elements, as well as the corresponding
energy eigenvalues $\omega_m$, are determined by solving the eigenvalue equations of motion
for each angular momentum and parity $J^{\pi}$
\begin{equation}
\left( \begin{array}{cc}
A & B \\ -B & - A 
\end{array} \right)
\left( \begin{array}{c}
X \\ Y \end{array} \right)
= \omega \left(
 \begin{array}{c}
X \\ Y \end{array} \right) ~.
\label{eq_rpa}
\end{equation}

The matrices $A$ and $B$ are (see e.g. \cite{Suh88})

\begin{eqnarray}
& & A^J_{pn,p'n'}  =  \\  \nonumber
& & \langle O | [ (c^{}_p c^{}_n )^{(JM) ^{\dagger}}, [ \hat{H},
(c^{\dagger}_{p'} c^{\dagger}_{n'} )^{(JM)}]] | O \rangle \\ \nonumber
 & = &  \delta_{pp'}\delta_{nn'} (E_p + E_n)  - \\
 \nonumber
&  & (u_p v_n u_{p'} v_{n'}+v_p u_n v_{p'} u_{n'} ) \times \\ \nonumber
& &  2 g_{ph} \langle p n^{-1},J | V | p' n'^{-1},J \rangle
\\ \nonumber
& & - (u_p u_n u_{p'} u_{n'}+v_p v_n v_{p'} v_{n'} ) \times \\ \nonumber
& &  2 g_{pp} \langle p n,J | V | p' n',J \rangle ~,
\label{eq:Adef}
\end{eqnarray}
and
\begin{eqnarray}
& & B^J_{pn,p'n'}   =   \\ \nonumber
& & \langle O | [ (c^{}_p c^{}_n )^{(J - M)} (-1)^M , [\hat{H}, 
(c^{}_{p'} c^{}_{n'} )^{(JM)}]] | O \rangle  \\
 \nonumber
& & =  - (u_p v_n v_{p'} u_{n'} + v_p u_n u_{p'} v_{n'} )  \\ \nonumber
& & \times 2 g_{ph} \langle p n^{-1} , J | V | p' n'^{-1}, J \rangle\\ \nonumber
& & + (u_p u_n v_{p'} v_{n'} + v_p v_n u_{p'} u_{n'} )  \\ \nonumber
&& \times 2 g_{pp} \langle p n , J | V | p' n', J \rangle ~,
\label{eq:Bdef}
\end{eqnarray}
where $E_p, E_n$ are the quasiparticle energies. 

The definitions, Eqs. (12) and  (13),  contain two renormalization adjustable parameters
$g_{ph}$ for the particle-hole interaction, and $g_{pp}$ for the particle-particle interaction. While $g_{ph} = 1.0$
is typically used, it is customary to adjust $g_{pp}$ so that the experimentally known half-life of the $2\nu\beta\beta$
decay is correctly reproduced \cite{us1}. But the particle-particle neutron-proton interaction governed by $g_{pp}$ actually 
contains two kinds of interaction matrix elements, isovector and isoscalar. Thus, to be consistent with the treatment
of the like particle pairing, one should separate the $T=1$ part from the $T=0$ part, i.e. replace
\begin{eqnarray}
& & g_{pp} \langle p n , J | V | p' n', J \rangle \rightarrow \\ \nonumber
& & g_{pp}^{T = 1} \langle p n , J, T=1 | V | p' n', J, T=1 \rangle \\ \nonumber 
& & + g_{pp}^{T=0} \langle p n , J, T=0 | V | p' n', J, T=0 \rangle ~,
\label{eq:sep}
\end{eqnarray}
and adjust the parameters $g_{pp}^{T = 1}$ and $g_{pp}^{T = 0}$ independently. To 
partially restore the isospin symmetry
and achieve that Eq.(\ref{eq:2nuF0}) is obeyed, it is sufficient to choose $g_{pp}^{T = 1} \sim g_{pair}$.
(That the coupling constant of the isovector proton-neutron particle-particle force should be close,
or identical, to the pairing strength constant, was recognized already in the early works on the QRPA
application to the $\beta\beta$ decay that used a schematic, $\delta$-force interaction, see Ref. \cite{VZ86}).

\begin{table*}[htb] 
\caption{   
Renormalization parameters of the pairing interaction, their average and the $T=1$ renormalization
constant $g_{pp}^{T=1}$ adjusted such that $M^{2\nu}_{F(cl)}$  and $M^{2\nu}_F$ vanish.
}\label{tab:pair}    
\centering 
\renewcommand{\arraystretch}{1.1}  
\begin{tabular}{lccccccccccc}
\hline \hline 
nucleus & NN pot. &  number &  & $d^{(i)}_{pp}$ & $d^{(i)}_{nn}$ & $d^{(f)}_{pp}$ & $d^{(f)}_{nn}$ & ~~$\bar{d}$~~ &
 & $g^{T=1}_{pp}$  \\ 
& & of s.p. lev & & & & & & & & \\ \hline
$^{48}$Ca & Argonne & 21 lev. & & --- & --- & 1.075 &  0.988 & 1.034 & & 1.031 \\
           & CD-Bonn & 21 lev.  & & --- & --- & 0.985 & 0.903 & 0.944  & & 0.944  \\
${^{76}Ge}$ & Argonne & 21 lev. & & 0.930 & 1.074 & 0.970 & 1.106 &  1.020 & & 1.038  \\
           & CD-Bonn & 21 lev. & & 0.863 & 0.983 & 0.899 & 1.013 &  0.940 & &  0.958  \\
${^{82}Se}$ & Argonne & 21 lev. & & 0.869 & 1.085 &  0.930 & 1.131 &  1.004 & &  1.032  \\
           & CD-Bonn & 21 lev. & & 0.808  & 0.995 &  0.864 & 1.038 &  0.926 & &  0.955 \\
${^{96}Zr}$ & Argonne & 21 lev. & & 0.923 & 0.768 &  1.000 & 0.962 &  0.913 & &  0.984 \\
            & CD-Bonn & 21 lev. & & 0.856 & 0.704 &  0.926 & 0.881 &  0.842 & & 0.907  \\
${^{100}Mo}$ & Argonne & 21 lev. & & 1.019 & 0.960 &  1.041 & 0.979 & 1.000 & & 1.008 \\
            & CD-Bonn & 21 lev. & & 0.946 & 0.883 &  0.966 & 0.900  & 0.924 & &  0.933 \\
${^{110}Pd}$ & Argonne & 21 lev. & & 1.000 & 0.975 & 1.025 & 0.945  & 0.986  & &  0.979   \\
            & CD-Bonn & 21 lev. & & 0.930 & 0.895  & 0.954 & 0.871  & 0.913  & &  0.908   \\
${^{116}Cd}$ & Argonne & 21 lev. & & 1.017  & 0.971 &  ---    & 0.919  & 0.969  & & 0.922  \\
            & CD-Bonn & 21 lev. & & 0.949  & 0.895 &  ---    & 0.847  & 0.897  & & 0.852   \\    
${^{124}Sn}$ & Argonne & 23 lev. & &  ---    & 1.001  &  0.929 & 1.000  &  0.977  & & 0.988  \\
            & CD-Bonn & 23 lev. & &  ---    & 0.918  &  0.860  & 0.917  & 0.898  & & 0.913  \\
${^{128}Te}$ & Argonne & 23 lev. & & 0.881 & 0.968  & 0.926 & 0.999  & 0.944   & & 0.988 \\
            & CD-Bonn & 23 lev. & & 0.816  & 0.889 & 0.857 & 0.918  & 0.870   & & 0.914  \\
${^{130}Te}$ & Argonne & 23 lev. & &  0.845 & 0.970  &  0.920 & 1.000  & 0.934   & & 0.989 \\
            & CD-Bonn & 23 lev. & &  0.783 & 0.891  &  0.852 & 0.918  & 0.861  & &  0.915  \\
${^{134}Xe}$ & Argonne & 23 lev. & & 0.851 & 0.912 &  0.917 & 0.963 & 0.911 & & 0.973  \\
           & CD-Bonn & 23 lev.  & & 0.790 & 0.840 &  0.850 & 0.887  & 0.842 & &  0.903 \\
${^{136}Xe}$ & Argonne & 23 lev. & & 0.782 &   ---   & 0.885 & 0.926  &  0.864 & &  0.950 \\
           & CD-Bonn  & 23 lev. & & 0.726 &  ---     & 0.821 & 0.853   & 0.800 & & 0.881 \\
\hline \hline  
\end{tabular}  
\end{table*}    

\section{Determination of the parameter $g_{pp}^{T=1}$}
\label{S:param}

When solving the BCS pairing equations, 
it is customary to slightly renormalize the strength of the pairing part
of the realistic nucleon-nucleon interaction so that
experimental pairing gaps are correctly reproduced. Thus, four adjusted parameters 
($d_{pp}^{(i)}, d_{pp}^{(f)}, d_{nn}^{(i)}, d_{nn}^{(f)}$) are introduced  
(see, e.g. \cite{us5,us1,us2,us3}) representing the adjustments needed
to describe the neutron and proton pairing gaps in the 
initial and final nuclei. The values of these parameters as well as their averages for selected
$\beta\beta$-decay candidate nuclei are displayed in Table \ref{tab:pair}.
(The Table entries are for two variants of the nucleon-nucleon interaction and one choice,
of large size (21/23 levels, oscillator shells $N = 0 - 5$
with the addition of the $i_{13/2}$ and $i_{11/2}$ for the nuclei 
heavier than $^{124}$Sn), of the single particle level scheme. The results for other choices are
not very different.) In several cases in  Table \ref{tab:pair} we encounter magic numbers of
neutrons or protons. In those cases the BCS treatment is inappropriate and hence the
corresponding entries are missing there. For the case of $^{48}$Ca we considered two variants.
In the listed one we assumed that there is no pairing in the doubly magic $^{48}$Ca.
In the other variant we assumed that the  values $\Delta_p$ = 2.18 MeV and $\Delta_n$ = 1.68
obtained from the usual odd-even mass difference with the five point formula represent 
the pairing gaps; the resulting $g_{pp}^{T=1}$ is rather similar to the values listed
in Table \ref{tab:pair}.

\begin{figure}[htbp]
\includegraphics[width=.50\textwidth,angle=0]{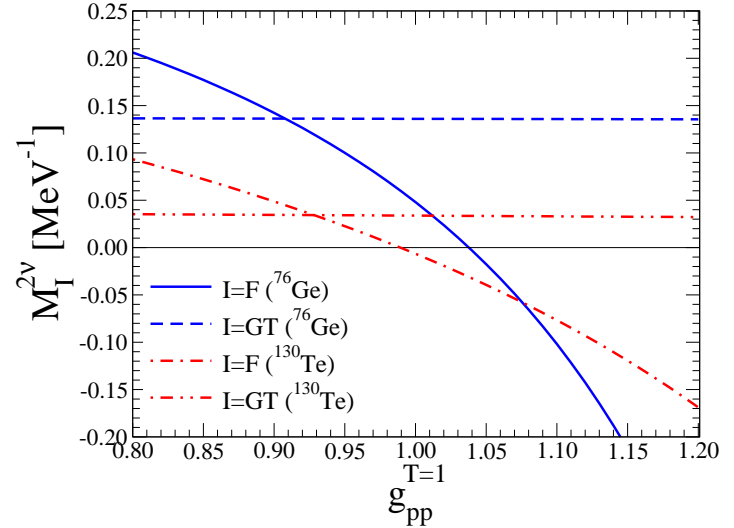}
\caption{(Color online) Dependence of the $2\nu\beta\beta$  matrix elements $M_F^{2\nu}$ and
$M_{GT}^{2\nu}$ on the isovector coupling constant $g_{pp}^{T=1}$. This example is for 
$^{76}$Ge and $^{130}$Te, the Argonne V18 potential, and the isoscalar coupling constants $g_{pp}^{T=0}$ = 0.750
and 0.783.
}
\label{fig:gT=1}
\end{figure}

The example in Fig. \ref{fig:gT=1} shows how the matrix elements $M_F^{2\nu}$ and
$M_{GT}^{2\nu}$ behave when the isovector coupling constant $g_{pp}^{T=1}$ is changed while
the isoscalar $g_{pp}^{T=0}$ is kept constant. As one can see, the Fermi matrix element $M_F^{2\nu}$
decreases and crosses zero, 
with increasing  $g_{pp}^{T=1}$, while the Gamow-Teller matrix element remains
constant. This is a typical case, and we can now choose $g_{pp}^{T=1}$ such that $M_{F-cl}^{2\nu}$
and hence also $M_F^{2\nu}$ vanish. Those values of $g_{pp}^{T=1}$ are shown in the last column 
of Table \ref{tab:pair}.

It follows from the entries in Table \ref{tab:pair} that 
the renormalization parameter $g_{pp}^{T = 1}$ that assures the
validity of the Eq. (\ref{eq:2nuF0}) is indeed very close to the average $\bar{d} \equiv g_{pair}$ 
of the pairing parameters $d_{ii}^{i, f}$.
In few rare cases, in particular in semi-magic nuclei,
the difference between  $g_{pp}^{T = 1}$ and $\bar{d}$ is $\sim$10\% (but not more). 
As shown in Ref. \cite{Vadim11} the ratio $g_{pp}^{T=1}/\bar{d}$ remains essentially
unchanged when the size of the single-particle basis is modified.

In this work, our choice is
to renormalize $g_{pp}^{T = 1}$ independently, but very close, to $\bar{d}$.
On the
other hand, it turns out that the $M^{2\nu}_{GT}$ depends sensitively only on  $g_{pp}^{T = 0}$, so that this
parameter can be still adjusted such that the half-life of the $2\nu\beta\beta$ decay is correctly reproduced, exactly
as done before. In fact, the previously used common value of $g_{pp}$, and the new parameter $g_{pp}^{T = 0}$, 
are essentially the same as we demonstrate in the next Section.


\section{Results and Discussion}

In the previous Section we explained how the parameter $g_{pp}^{T = 1}$ is determined. The determination of the
other renormalization parameter $g_{pp}^{T = 0}$ is analogous and follows the suggestion made long time ago in 
Ref. \cite{us1} as already stated. We fit $g_{pp}^{T = 0}$ from the requirement that the calculated values of the full $2\nu\beta\beta$
matrix elements $M_{GT}^{2\nu}$ agrees with their experimental values. 
For most nuclei in Table I the half-lives $T^{2\nu}_{1/2}$
have been measured; we use the recommended values in Ref.\cite{Barab} 
plus the $^{136}$Xe half-life of Refs. \cite{EXO,ZEN}. But for several nuclei in that Table, and in the Tables that follow,
the half-life remains unknown. In those cases we proceed as follows: for $^{110}$Pd we use the estimate
of Ref. \cite{Pd110} that uses the single-state dominance assumption, for $^{124}$Sn and $^{134}$Xe we use an interval
of possible $M^{2\nu}_{GT}$ values $0 \le M^{2\nu}_{GT} \le  0.2(0.1)$MeV$^{-1}$ for $^{124}$Sn($^{134}$Xe) , respectively. 
For these two nuclei we show in Tables II - IV  the results with $g_A$ = 1.0 for the upper limit of $M^{2\nu}_{GT}$,
and with $g_A = 1.27$ for the lower limit $M^{2\nu}_{GT} = 0$. Our results for these two nuclei 
therefore reflect our incomplete knowledge of the corresponding $2\nu$
half-life.

\begin{figure}[htb]
\includegraphics[width=.50\textwidth,angle=0]{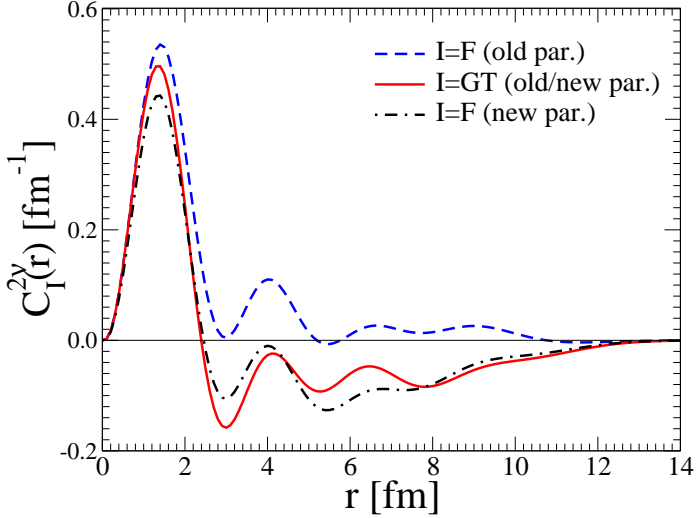}
\caption{ (Color online) Functions $C^{2\nu}_F(r)$ with old and new parametrization and, for comparison,
the function $C^{2\nu}_{GT}(r)$ (scaled by 1/3) for clarity) is also shown. This is the case of $^{76}$Ge. }
\label{fig:newold}
\end{figure}

Before presenting the results for the $0\nu\beta\beta$ nuclear matrix elements, several comments are in order.
Since the main effect considered here is the change in $M^{2\nu}_F$ and the associated change in $M^{0\nu}_F$,
lets analyze these changes using the radial dependence of  $M^{2\nu}_{F-cl}$ explained in Ref.\cite{us5}. 
In Fig. \ref{fig:newold} the functions $C^{2\nu}_{F-cl}(r)$ with the old and new parametrization of $g_{pp}$ are 
plotted, together with the function $C^{2\nu}_{GT-cl}(r)$ (scaled by 1/3 for clarity). As one can see, with the
new   $g_{pp}^{T = 1}$ the tail of $C^{2\nu}_{F-cl}(r)$ becomes more negative and therefore its integral vanishes,
as required. Let us remind ourselves that
\begin{equation}
M^{2\nu}_{F-cl} = \int_0^{\infty} C^{2\nu}_{F-cl}(r) dr ~,
\end{equation}
and in analogy
\begin{equation}
M^{2\nu}_{GT-cl} = \int_0^{\infty} C^{2\nu}_{GT-cl}(r) dr ~.
\end{equation}

Another comment concerns the fact that, as we will see, with the new parametrization $\chi_F \approx -(0.3 - 0.4)$,
or more precisely $\chi_F \approx -1/3$. Somewhat similar conclusion is obtained in the shell model \cite{ISM}
where the isospin is conserved by definition.
The explanation is based on the fact that the ground states of even-even nuclei consist dominantly
of the $J^{\pi} = 0^+, T = 1$ Cooper pairs that, in turn, are mostly in the $S=0, L=0$ state. Since such
states are eigenstates of the operator $\vec{\sigma}_1 \cdot \vec{\sigma}_2$ with eigenvalue -3, our conclusion
simply follows. 

\begin{figure}[htb]
\includegraphics[width=.55\textwidth,angle=0]{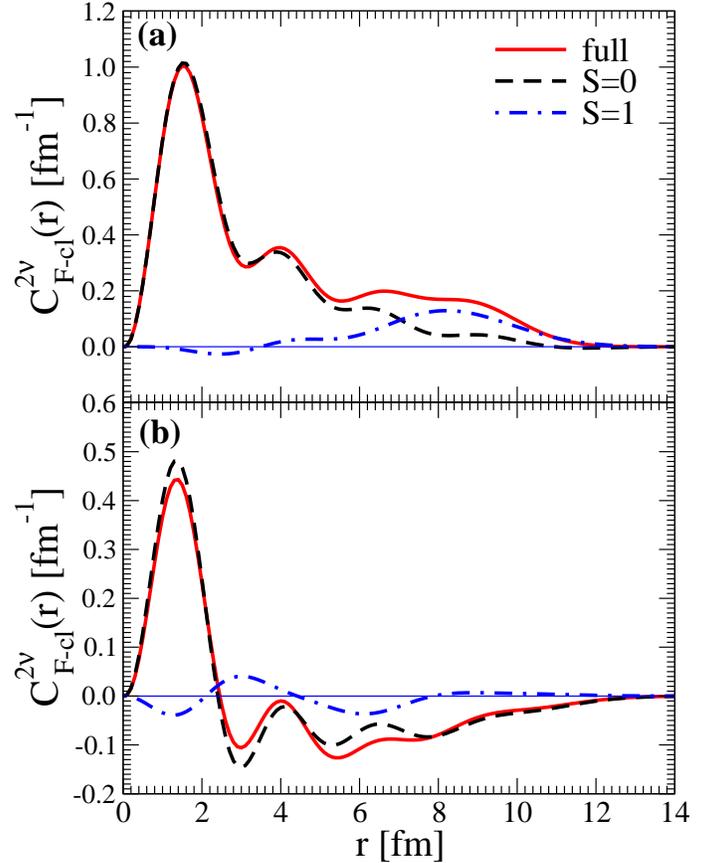}
\caption{ (Color online) In the upper panel (a) the function $C^{2\nu}_{F-cl}(r)$ is shown for the pure
pairing case, i.e. $g_{pp}^{T=0} = g_{pp}^{T=1} = g_{ph} = 0.0$, separated into the $S=0$
and $S = 1$ components. In the lower panel (b) the
 function $C^{2\nu}_{F-cl}(r)$ is shown for $g_{pp}^{T=1}$ = 1.038 and
 $g_{pp}^{T=0} = 0.750$ again separated into its $S=0$ and $S=1$ parts.
The sum function is also displayed. The dominance of the $S=0$ component is clearly
visible in the upper panel. In the lower panel the two components when integrated over $r$
are, naturally, equal and opposite. The $S=0$ part, however, clearly is considerably
larger in absolute value than the $S = 1$ part, at all $r$ values.
 This is the case of $^{76}$Ge.
 }
\label{fig:s01}
\end{figure}

In Fig. \ref{fig:s01} we show examples of the decomposition of the function
 $C^{2\nu}_{F-cl}(r)$ into their $S=0$ and $S=1$ components. These are rather typical cases.
 The dominance of the $S = 0$ component in the pure pairing case (the upper panel in Fig. \ref{fig:s01}) is
 easily understood. However, that feature is still present in the realistic case with $g_{pp} \ne 0.0$, hence our finding that,
 usually, $\chi_F \approx -1/3$. 
 
 For both modes, $0\nu\beta\beta$ and $2\nu\beta\beta$, we can find relations 
 between the Fermi and Gamow-Teller parts and their $S = 0$ and $S = 1$ components. These relations are exact in
 the closure approximation and when the higher order weak currents ( and thus the tensor part $M^{0\nu}_T$ ) are
 neglected and the nucleon form factors have the same cut-off values for the vector and axial vector parts.
 In addition, since the neutrino potentials indirectly depend on the assumed averaged energy, these $\bar{E}$
 values must be chosen to be the same for the Fermi and Gamow-Teller matrix elements.
 Using the properties of the $\vec{\sigma_1} \cdot \vec{\sigma_2}$ operator and that $M^{2\nu}_F = 0$ with
 our new parametrization we find that only one of the four components is independent and
 \begin{eqnarray}
 & & M^{2\nu}_{cl-F}(S=0) = -M^{2\nu}_{cl-F}(S=1) = 
 \\ \nonumber
 & &  = -M^{2\nu}_{cl-GT}(S=1) = - M^{2\nu}_{cl-GT}(S=0)/3 ~.
 \end{eqnarray}
 
 For the $0\nu$ mode, however, $M^{0\nu}_F \ne 0$ and hence the above relations must be modified:
 \begin{eqnarray}
 & & M^{0\nu}_F(S=0) =  M^{0\nu}_F - M^{0\nu}_F(S=1) = 
 \\ \nonumber
 & &  = M^{0\nu}_F - M^{0\nu}_{GT}(S=1) = - M^{0\nu}_{GT}(S=0)/3 ~,
 \\ \nonumber
 & & M^{0\nu}_{GT} = M^{0\nu}_F - 4 M^{0\nu}_F(S=0) ~.
 \end{eqnarray}
 Two components are independent in this case. In realistic case these relations are not exact, but still 
 valid in a reasonable approximation.
 
 We will return to the discussion of the $\chi_F$ values obtained by different approximate
 methods in the next Section.

In tables \ref{tab:res1} and \ref{tab:res2} we compare the resulting matrix elements $M^{2\nu}_F$, $M^{2\nu}_{GT}$
and $M^{0\nu}$ with its components evaluated using the old parametrization ($g_{pp}^{T=1} = g_{pp}^{T=0} \equiv g_{pp}$) 
with the
new results, where $g_{pp}^{T=1} \approx g_{pair}$ and where $g_{pp}^{T=0}$ is fitted to the known
experimental values of $M^{2\nu}_{GT}$. The calculations were performed for the unquenched
value $g_A = 1.27$ as well as for $g_A = 1.0$. The quantities $M'^{0\nu} = M^{0\nu} \times (g_A / 1.27)^2$
are also shown, as well as $\chi_F$. Calculations in both tables were performed within the standard
QRPA with all the usual ingredients, i.e. including the higher order weak currents, nucleon form factors,
and the short-range correlation treatment of Ref. \cite{Sim09}.

Lets explain briefly again (for more details see Ref. \cite{us5}) how the quenching is taken
into account with our method. Since we adjust the isoscalar particle-particle renormalization constant
$g_{pp}^{T = 0}$ in such a way that the experimental half-life of the $2\nu\beta\beta$ is correctly reproduced, 
by changing the effective value of the axial current coupling constant $g_A$ we are forced
to change also the parameter  $g_{pp}^{T = 0}$, albeit only slightly. Those changes are visible
in the third columns of Tables II and III.  Since with smaller $g_A$ the parameter $g_{pp}^{T = 0}$
slightly decreases, the corresponding $M^{0\nu}_{GT}$ matrix element increases. However,
the $0\nu\beta\beta$ decay rate, proportional to the $(M'^{0\nu})^2$, naturally, decreases.

In that context it is worthwhile to point out another feature of the new parametrization. The
Fermi matrix element $M^{0\nu}_F$ is associated with the weak vector current, and as
such should not be affected by the axial current quenching. With the old parametrization,
with a single $g_{pp}$, that was not quite true, as seen in Tables II and III. However, with
the new parametrization where isospin symmetry is partially restored, the $M^{0\nu}_F$
becomes independent of the effective value of $g_A$ as it should be. 
(The tiny changes in Tables II and III
are round-off errors.)

From the tables one can see that the new parametrization, leading to $M^{2\nu}_F = 0.0$, leads
to a substantial reduction of the $M^{0\nu}_F$ component of $M^{0\nu}$ and overall
$\sim$ 10-20\% reduction of the final $M^{0\nu}$ nuclear matrix elements. It is encouraging
that both variants of the $M^{0\nu}$ matrix elements for $^{48}$Ca are now rather
close to the results of nuclear shell model evaluation. (With $g_A$ = 1.27 our $M^{0\nu}$
values are 0.54 in the listed case and 0.71 in the variant where the even-odd mass
differences are treated as arising from pairing, both with the
Argonne V18 potential and 0.59 (0.77) with the CD-Bonn potential , while the shell model values are
0.59 in Ref. \cite{Cau08} and 0.82(0.90) for the Argonne V18 (CD Bonn) potential in
Ref.\cite{Horoi}.) Note that only in the case of $^{48}$Ca the full oscillator
$pf$ shell is included and hence the Ikeda sum rule is fulfilled in the nuclear shell model treatment. 
We are, naturally, well aware of the fact that to apply QRPA in the case of $^{48}$Ca is questionable;
our results should be treated with that in mind.

\begin{figure}[htb]
\includegraphics[width=.53\textwidth,angle=0]{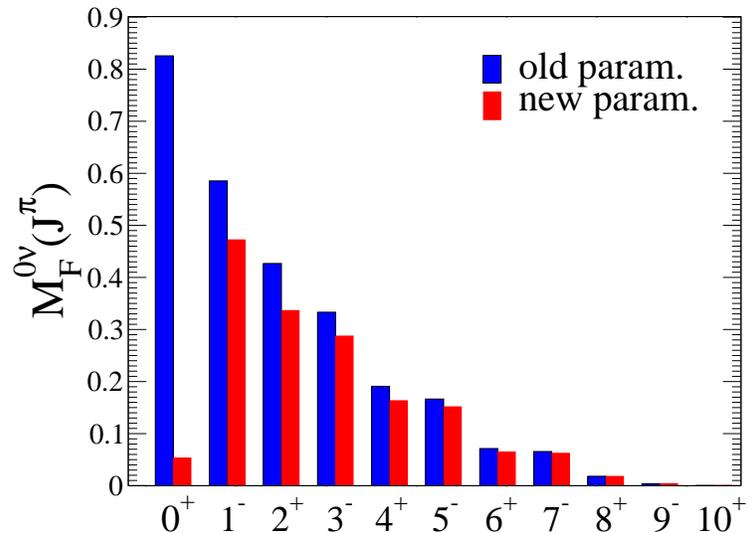}
\caption{ (Color online) Multipole decomposition of the matrix element $M^{0\nu}_F$. The results with the old and
new parametrizations are compared. Note the dominant effect for the $0^+$ multipole, and
the relatively small effects for the other multipoles. This is the case of $^{76}$Ge. }
\label{fig:fmult}
\end{figure}

Finally, in order to better visualize effect of the new parametrization of the particle-particle interaction,
we show in Fig. \ref{fig:fmult} an example of the multipole decomposition of the matrix element $M^{0\nu}_F$.
One can see there that the contribution of the intermediate multipole $0^+$ is drastically reduced with
our choice of $g_{pp}^{T=1}$, while all the other multipoles are affected only slightly or not at all.
This is, in some sense, analogous to the situation with $M^{0\nu}_{GT}$ where the parameter $g_{pp}^{T=0}$
affects mostly the intermediate $1^+$ states, while all the other multipolarities are affected much less.

\begin{table*}[htb] 
\caption{ Nuclear matrix elements for both $\beta\beta$ decay modes with the old parametrization
($g_{pp}^{T = 0} = g_{pp}^{T = 1} \equiv g_{pp}$)  are compared to those with the new one, $g_{pp}^{T = 0} \ne g_{pp}^{T = 1}$.
The adopted values of the parameter $g_{pp}^{T=0}$ are also shown. The results for two values of the axial
coupling constant $g_A$ are displayed; the quenched value $g_A = 1.0$ and the standard value $g_A = 1.27$. 
The G-matrix elements of realistic Argonne V18 potential nucleon-nucleon potential are considered.
The nuclear radius $R=r_0 A^{1/3}$ with $r_0$ = 1.2fm is used.
}
\label{tab:res1}    
\centering 
\renewcommand{\arraystretch}{1.1}  
\begin{tabular}{lcccccccccccccccc}
\hline \hline 
nucleus & $g_A$ & $g_{pp}^{T=0}$ & param. &  & \multicolumn{2}{c}{$2\nu\beta\beta$-decay NMEs} & &
\multicolumn{5}{c}{$0\nu\beta\beta$-decay NMEs}  & & $\chi_F$ \\ \cline{6-7} \cline{9-13} 
        &  & &  & & $M^{2\nu}_F$ & $M^{2\nu}_{GT}$ & & $M^{0\nu}_F$ & $M^{0\nu}_{GT}$ & $M^{0\nu}_T$ & ${M}^{0\nu}_{}$  
& ${M'}^{0\nu}_{}$ & & \\ \hline\hline
 ${^{48}Ca}$ & 1.000 & 0.771  & old   & & 0.331  & 0.0736  & & -0.794 & 0.642  & -0.164  & 1.272  & 0.790 & & -1.24 \\    
              &      & 0.770 & new   & & 0.00  & 0.0736  & &  -0.268 & 0.639   & -0.161 & 0.745  & 0.463 & & -0.42 \\
             & 1.27 & 0.776  & old   & & 0.327  & 0.0457 & &  -0.788  & 0.526  & -0.151 & 0.864 & 0.864  & & -1.50  \\   
              &      & 0.775 & new   & & 0.00  & 0.0457  & &  -0.268 &  0.523  & -0.149 & 0.541 & 0.541 & &  -0.51  \\ 
 ${^{76}Ge}$ & 1.00  & 0.728  & old   & & 0.240    & 0.220 & & -2.688 & 5.276    & -0.606  & 7.357  & 4.569 & & -0.51 \\
             &      &  0.728  & new   & & 0.00    & 0.220 & & -1.612 & 5.236    & -0.591  & 6.258  & 3.886 & & -0.31  \\
             & 1.27 &  0.750  & old   & & 0.231   & 0.137 & & -2.632 & 4.753    & -0.575  & 5.812  & 5.812 & &  -0.55  \\
             &      &  0.750  & new   & & 0.00    & 0.137 & & -1.615 & 4.715    & -0.561  & 5.157  & 5.157 & & -0.34  \\
 ${^{82}Se}$ & 1.00  &  0.751  & old   & & 0.180    & 0.153 & & -2.394 & 4.614    & -0.557  & 6.452  & 4.007 & & -0.52  \\
             &      &  0.751   & new   & & 0.00    & 0.153 & & -1.529 & 4.586    & -0.545  & 5.571  & 3.460 & & -0.33  \\
             & 1.27 &  0.766   & old   & & 0.175   & 0.095 & & -2.359 & 4.233    & -0.527  & 5.171  & 5.171 & & -0.56  \\
             &      &  0.766   & new   & & 0.00    & 0.095 & & -1.531 & 4.207    & -0.516  & 4.642  & 4.642 & & -0.36  \\
 ${^{96}Zr}$ & 1.00  &  0.806  & old   & & 0.063   & 0.145 & & -1.547 & 2.825    & -0.414  & 3.958  & 2.458 & & -0.55  \\
             &      &  0.817   & new   & & 0.00    & 0.145 & & -1.214 & 2.667    & -0.411  & 3.469  & 2.154 & & -0.45  \\
            & 1.27  &  0.824  & old   & & 0.058   & 0.090 & & -1.518 & 2.466    & -0.391  & 3.018  & 3.018  & & -0.62 \\
           &      &   0.830   & new   & & 0.00    & 0.090 & & -1.215 & 2.349    & -0.387  & 2.717  & 2.717 & & -0.52  \\
 ${^{100}Mo}$ & 1.00 & 0.841   & old   & & 0.100   & 0.373 & & -2.757 & 5.166    & -0.683  & 7.240  & 4.496 & & -0.53  \\
           &      &   0.840   & new   & & 0.00    & 0.373 & & -2.250 & 5.162    & -0.673  & 6.739  & 4.185 & & -0.44  \\
           & 1.27 &   0.848   & old   & & 0.097   & 0.232 & & -2.738 & 4.640    & -0.645  & 5.696  & 5.696 & & -0.59  \\
           &      &   0.847  & new   & & 0.00    & 0.232 & & -2.251 & 4.639    & -0.635  & 5.402  & 5.402 & & -0.49  \\
${^{110}Pd}$ & 1.00 & 0.785 & old   & & 0.081 & 0.423  & &  -2.668 & 5.609    & -0.585  & 7.692  & 4.777 & & -0.48  \\
             &      & 0.783 & new   & & 0.00  & 0.423 & &  -2.182  & 5.614    & -0.574  & 7.222  & 4.485 & & -0.39 \\
             & 1.27 & 0.805 & old  & &  0.075 & 0.263  & & -2.626 & 4.949     & -0.558  & 6.021 & 6.021  & & -0.53 \\
             &      & 0.803 & new   & & 0.00  & 0.263 & &  -2.184 & 4.954    & -0.549  & 5.762 & 5.762  & & -0.44   \\ 
${^{116}Cd}$ & 1.00 & 0.870  & old   & & 0.008  & 0.206 & & -1.633 & 3.663  &  -0.277  & 5.019 & 3.117 & & -0.45 \\
             &      & 0.870  & new   & & 0.00  & 0.206 & &  -1.583 & 3.661   &  -0.275 &  4.969 & 3.086 & & -0.43 \\
             & 1.27 & 0.900  & old   & & 0.004 & 0.128 & &  -1.607 & 3.319   &  -0.264 &  4.053 & 4.053  & & -0.48 \\
             &      & 0.900  & new   & & 0.00  & 0.128  & &  -1.586  & 3.318  & -0.263 &  4.040 & 4.040  & & -0.48  \\ 
 ${^{124}Sn}$ &  1.00    & 0.628  & old   & & 0.132  & 0.20 & & -1.779  & 3.860  & -0.344 &  5.295 & 3.288 & &  -0.46  \\
             &      & 0.626  & new   & & 0.00   & 0.20  & & -0.984 & 3.859  & -0.338 &  4.504  & 2.797 & & -0.25  \\
             & 1.27     & 0.785  & old   & & 0.086  & 0.00  & & -1.473 & 2.308  &  -0.361 &  2.861  & 2.861 & & -0.64   \\
             &      & 0.785  & new   & & 0.00   & 0.00  & & -0.988 & 2.302  &  -0.358 &  2.558  & 2.558  & & -0.43 \\
 ${^{128}Te}$ & 1.00 & 0.770  & old   & & 0.133 & 0.0776 & & -2.540 & 4.453  & -0.642 & 6.351  & 3.944 & &  -0.57  \\
             &      & 0.769  & new   & & 0.00  & 0.0776 & & -1.750 & 4.436  & -0.634  & 5.552  & 3.445  & & -0.39   \\
             & 1.27 & 0.780  & old   & & 0.128 &  0.0481 & & -2.508 & 4.092  & -0.608 &  5.042  & 5.042  & & -0.61  \\
             &      & 0.779  & new   & & 0.00  &  0.0481 & & -1.751 & 4.076  & -0.601 &  4.563  & 4.563  & & -0.43  \\
 ${^{130}Te}$ & 1.00 & 0.775 & old   & & 0.103 & 0.0545  & & -2.232 & 3.796  & -0.588 & 5.439 & 3.378  & &  -0.59   \\
             &      & 0.774 & new   & & 0.00  & 0.0545  & & -1.545 & 3.778  & -0.582 & 4.742 & 2.945   & & -0.41   \\
             & 1.27 & 0.784  & old   & & 0.100 & 0.0339  & & -2.206 & 3.493 & -0.556  & 4.306 & 4.306  & &-0.63   \\
             &      & 0.783  & new   & & 0.00  & 0.0339  & & -1.546 & 3.478 & -0.550  & 3.888 & 3.888  & & -0.44  \\
 ${^{134}Xe}$ &  1.00    & 0.739 & old   & & 0.111  & 0.10   & & -2.247 & 4.108  & -0.537  & 5.819 & 3.613  & & -0.55 \\   
             &      & 0.738  & new   & & 0.00  & 0.10   & & -1.501  & 4.091 & -0.530  & 5.071 & 3.149 & & -0.37 \\
             & 1.27      & 0.787  & old   & & 0.092 & 0.00   & & -2.112  & 3.256  &  -0.521 & 4.045 & 4.045 & & -0.65  \\
              &      & 0.787   & new   & & 0.00  & 0.00   & & -1.513 & 3.241  &  -0.517 & 3.664 & 3.664 & & -0.47 \\                               
  ${^{136}Xe}$ & 1.00 & 0.730 & old   & & 0.0652  & 0.0313 & & -1.228 & 2.149   & -0.299 & 3.078  & 1.911 & & -0.57  \\
           &        & 0.730   & new   & & 0.00     & 0.0313 & & -0.806 & 2.138   &  -0.297 & 2.646 &  1.643 & & -0.38  \\
           & 1.27 &  0.740   & old   & & 0.0627   & 0.0194 & &  -1.211 & 1.968  & -0.283  & 2.437 & 2.437 & & -0.62  \\
           &      &  0.740    & new   & & 0.00     & 0.0194 & &  -0.806 & 1.959  & -0.282  & 2.177 & 2.177  & & -0.41 \\
\hline \hline  
\end{tabular}  
\end{table*}    

\begin{table*}[htb] 
\caption{ The same as in Table \ref{tab:res1}  for realistic charge-dependent Bonn potential (CD-Bonn).
}
\label{tab:res2}    
\centering 
\renewcommand{\arraystretch}{1.1}  
\begin{tabular}{lcccccccccccccccc}
\hline \hline 
nucleus & $g_A$ & $g_{pp}^{T=0}$ & param. &  & \multicolumn{2}{c}{$2\nu\beta\beta$-decay NMEs} & &
\multicolumn{5}{c}{$0\nu\beta\beta$-decay NMEs}  & & $\chi_F$ \\ \cline{6-7} \cline{9-13} 
        &  & & & & $M^{2\nu}_F$ & $M^{2\nu}_{GT}$ & & $M^{0\nu}_F$ & $M^{0\nu}_{GT}$ & $M^{0\nu}_T$ & ${M}^{0\nu}_{}$  
& ${M'}^{0\nu}_{}$ & &  \\ \hline\hline
${^{48}Ca}$ & 1.000 & 0.703  & old   & & 0.321  & 0.0736  & & -0.795 & 0.678  & -0.151  & 1.322  & 0.821 & & -1.17 \\
             &      & 0.703  & new   & & 0.00  & 0.0736  & &  -0.284 & 0.675   & -0.149 & 0.810  & 0.503 & & -0.42 \\
             & 1.27 & 0.708 & old   & & 0.318  & 0.0457 & &  -0.789  & 0.558  & -0.139 & 0.909 & 0.909  & & -1.41  \\
             &      & 0.708  & new   & & 0.00  & 0.0457  & &  -0.284 &  0.555  & -0.137 & 0.594 & 0.594 & &  -0.51  \\
 ${^{76}Ge}$ & 1.00 & 0.660  & old   & & 0.188   & 0.220 & & -2.794 & 5.644    & -0.551  & 7.886  & 4.897 & & -0.49 \\
            &       & 0.660  & new   & & 0.00    & 0.220 & & -1.711 & 5.607    & -0.537  & 6.781  & 4.211 & & -0.30 \\
            & 1.27 & 0.681  & old   & & 0.220   & 0.137 & & -2.735 & 5.052    & -0.522  & 6.228  & 6.228  & & -0.54 \\
            &      & 0.681   & new   & & 0.00    & 0.137 & & -1.713 & 5.018    & -0.510  & 5.571  & 5.571 & & -0.34 \\
  ${^{82}Se}$ & 1.00 & 0.683  & old   & & 0.172   & 0.156 & & -2.495 & 4.939    & -0.508  & 6.926  & 4.301 & & -0.51 \\
            &       &  0.683  & new   & & 0.00    & 0.156 & & -1.616 & 4.913    & -0.497  & 6.032  & 3.746 & & -0.33  \\
            & 1.27  &  0.698  & old   & & 0.166   & 0.095 & & -2.459 & 4.508    & -0.480  & 5.555  & 5.555 & & -0.55 \\
            &       &  0.698  & new   & & 0.00    & 0.095 & & -1.618 & 4.484    & -0.470  & 5.018  & 5.018 & & -0.36 \\
 ${^{96}Zr}$  & 1.00 &  0.735 & old   & & 0.060   & 0.145 & & -1.614 & 3.035    & -0.373  & 4.276  & 2.655 & & -0.53 \\
             &      & 0.747   & new   & & 0.00    & 0.145 & & -1.277 & 2.861    & -0.370  & 3.769  & 2.341 & & -0.45  \\
             & 1.27 & 0.753   & old   & & 0.055   & 0.090 & & -1.583 & 2.640    & -0.351  & 3.271  & 3.271 & & -0.60 \\
             &      & 0.760   & new   & & 0.00    & 0.090 & & -1.278 & 2.511    & -0.347  & 2.957  & 2.957  & & -0.51  \\
  ${^{100}Mo}$  & 1.00 & 0.770  & old   & & 0.097 &  0.373 & &  -2.883 & 5.532  & -0.615  & 7.800 & 4.8435 & & -0.52 \\
              &      & 0.769  & new   & & 0.00  &  0.373 & &  -2.366 & 5.526  & -0.606 & 7.287 & 4.525 & & -0.43  \\
              & 1.27 & 0.778  & old   & & 0.093 &  0.232 & & -2.863 & 4.950  &  -0.580 & 6.148  & 6.148 & & -0.58 \\
              &      & 0.776  & new   & & 0.00  &  0.232 & & -2.367 & 4.950   & -0.571 & 5.850  & 5.850 & & -0.48   \\
  ${^{110}Pd}$ & 1.00 & 0.715 & old   & & 0.081  & 0.423  & & -2.799  & 6.046   & -0.531  & 8.314  & 5.163 & & -0.46  \\
             &      & 0.713   & new   & & 0.00      & 0.423 & & -2.288 & 6.052   & -0.521  & 7.820  & 4.856 & & -0.38  \\
             & 1.27 & 0.734 & old   & &  0.075 & 0.263 & &  -2.755 & 5.324   &  -0.506  & 6.529  & 6.529  & & -0.52 \\
             &      & 0.732  & new   & & 0.00  & 0.263 & & -2.290 & 5.330   &  -0.497   & 6.255  & 6.255  & & -0.43  \\
 ${^{116}Cd}$ & 1.00 & 0.785 & old   & & 0.010  & 0.206  & & -1.707 & 3.942  &  -0.253  & 5.396 & 3.351 & & -0.43 \\
             &      & 0.784 & new   & & 0.00  & 0.206  & &  -1.639 & 3.940  & -0.251 &  5.328 & 3.308  & & -0.42 \\
             & 1.27 & 0.815 & old   & & 0.006 & 0.128  & &  -1.680 & 3.564  & -0.241 & 4.367  & 4.367   & & -0.40\\
             &      & 0.814 & new   & & 0.00 & 0.128  & &  -1.642 & 3.563   & -0.240  & 4.343 & 4.343  & &  -0.46 \\
 ${^{124}Sn}$ &  1.00    & 0.557 & old   & &  0.127 & 0.200  & & -1.871 & 4.208  & -0.311  & 5.768 & 3.582  & & -0.44 \\
             &      & 0.555 & new   & &  0.00  & 0.200  & & -1.057 &  4.206 & -0.305  & 4.958 & 3.079  & & -0.25 \\
             &  1.27    & 0.708  & old   & & 0.085  & 0.00   & & -1.569  & 2.579  & -0.324  & 3.230  & 3.230  & & -0.61\\
             &      & 0.707  & new   & &  0.00  & 0.00    & & -1.062 & 2.575  & -0.321  & 2.913  & 2.913  & & -0.41 \\
 ${^{128}Te}$ & 1.00 & 0.694 & old   & & 0.130 & 0.0776 & & -2.673 & 4.902  & -0.580 & 6.996 & 4.344  & & -0.54 \\
             &      & 0.693 & new   & & 0.00  & 0.0776 & &  -1.850 & 4.887 & -0.572 & 6.164 & 3.828  & & -0.38 \\
             & 1.27 & 0.704 & old   & & 0.125 & 0.0481 & & -2.641 & 5.582  & -0.549 & 5.582 & 5.582  & & -0.47  \\
             &      & 0.703 & new   & & 0.00 & 0.0481 & & -1.851  & 4.476  & -0.542 & 5.084 & 5.084  & & -0.41 \\
 ${^{130}Te}$ & 1.00 & 0.698 & old   & & 0.102 & 0.0545  & & -2.354 & 4.213 & -0.531 & 6.036 & 3.748   & & -0.56   \\
             &      & 0.697 & new   & & 0.00  & 0.0545  & &  -1.637 & 4.198 & -0.525 & 5.310 & 3.297   & &  -0.39  \\
             & 1.27 & 0.707 & old   & & 0.098 & 0.0339  & & -2.328 &  3.867  & -0.502  & 4.810 & 4.810  & & -0.60  \\
             &      & 0.706 & new   & & 0.00 & 0.0339  & &  -1.637 &  3.852  & -0.496 &  4.373 & 4.373  & & -0.42  \\
   ${^{134}Xe}$ &  1.00    & 0.664 & old   & & 0.110 &  0.10 & & -2.268 & 4.532  & -0.486  & 6.414  & 3.983 & & -0.50  \\
             &      & 0.663 & new   & & 0.00  &  0.10 & & -1.595 & 4.516  & -0.479  & 5.632  & 3.497  & & -0.35  \\
             &   1.27   & 0.712 & old   & & 0.092 &  0.00 & & -2.231 & 3.608  & -0.472  & 4.522  & 4.522  & & -0.62    \\
             &      & 0.712 & new   & & 0.00   &  0.00 & & -1.599 & 3.593 & -0.466  & 4.119  & 4.119  & & -0.44     \\         
  ${^{136}Xe}$  & 1.00 &  0.657 & old   & & 0.0643   & 0.0131 & &  -1.300 & 2.395  & -0.269  & 3.426  & 2.127 & & -0.54\\
              &      & 0.657  & new   & & 0.00     & 0.0313 & &  -0.858 & 2.385  & -0.268  & 2.975  & 1.847 & & -0.36 \\
              & 1.27 & 0.667 & old   & & 0.0619   & 0.0194 & &  -1.282 & 2.190  & -0.255  & 2.735  & 2.735 & & -0.59 \\
              &      & 0.667  & new   & & 0.00     & 0.0194 & &  -0.858  & 2.181 & -0.254  & 2.460  & 2.460 & & -0.39  \\
\hline \hline  
\end{tabular}  
\end{table*}    

\begin{figure}[htb]
\includegraphics[width=.50\textwidth,angle=0]{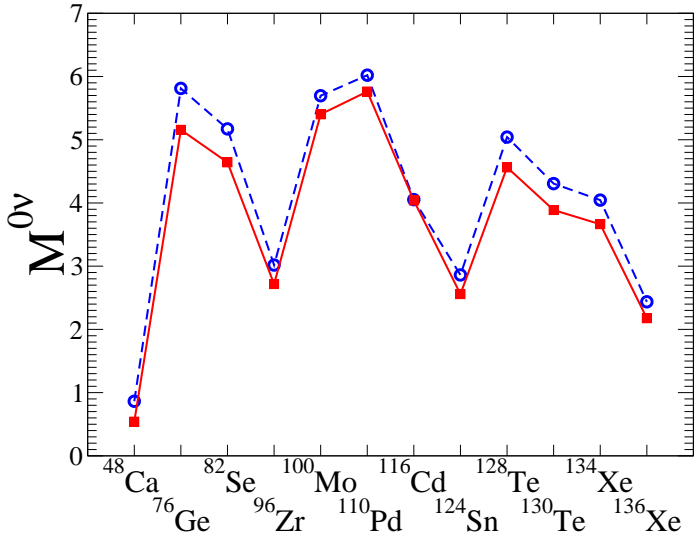}
\caption{(Color online) Nuclear matrix elements $M^{0\nu}$ evaluated with the new parametrization developed in this
work (filled squares) compared with the old method ($g_{pp}^{T=1} = g_{pp}^{T=0} \equiv g_{pp}$) 
(empty circles). This is
QRPA with $g_A$ = 1.27 and large size single particle level scheme, as in Table I,
evaluation using the Argonne V18 potential. 
 }
\label{fig:me0n}
\end{figure}

We compare in Fig. \ref{fig:me0n} the $M^{0\nu}$ matrix elements for all
considered nuclei evaluated with the old and new parametrizations of $g_{pp}$.
The smaller values of $M^{0\nu}$ in $^{48}$Ca, $^{166}$Cd, $^{124}$Sn, $^{136}$Xe
and to some extent also in $^{96}$Zr are related to the magic or semimagic
nucleon number in these nuclei, and thus to the reduced pairing correlations in them. 

 \section{Comparison of the $\chi_F$ values evaluated by different methods}
 
 As we argued in this work, the result of the new parametrization of the particle-particle
 interaction, that partially restores the isospin symmetry and leads to the correct
 $M^{2\nu}_F$ = 0 value, is the reduction of the Fermi part $M^{0\nu}_F$ of the
 $0\nu\beta\beta$ nuclear matrix element. At the same time, the largest component
 of that matrix element, $M^{0\nu}_{GT}$, remains essentially unaffected. One can see that 
 most clearly by considering the quantity $\chi_F$, the ratio $M^{0\nu}_F/M^{0\nu}_{GT}$.
 
 In Table \ref{tab:chif} we compare the $\chi_F$ values obtained with different methods.
 (Analogous table, naturally without our new results, appears in Ref. \cite{barea13} in their Table VII. 
 However, as we already mentioned, their definition of $\chi_F$ contains an extra factor $(g_V/g_A)^2.$)
 One can see in the Table \ref{tab:chif} that in the nuclear shell model, and in our QRPA calculation
 with the new parametrization of $g_{pp}$, the $\chi_F$ values are substantially smaller than in
 the previous approaches. (In IBM-2 the $\chi_F$ are very small when neutrons and protons are
 in different shells. That is an artifact of the model where only one shell in each system is included.)
 
 In the shell model, and in our new QRPA calculations, the $\chi_F$ values are relatively close
 to -1/3, the value one would obtain in pure $S=0$ states. However, in the shell model the
 $\chi_F$ values are systematically smaller than in our version of QRPA.
 Why this is so remains to be understood.
 (To be really precise, $\chi_F = -1/3$ would arise for pure $S  = 0$ when the higher order
 terms in the weak current are absent, when in the nucleon form factor the cut-off parameters
 for the vector and axial vector currents are the same and the average energies $\bar{E}$
 are the chosen to be the same in both neutrino potentials.) As we pointed out before, while the
 $S=0$ component is large, the other parts, in particular $S=1$, are clearly present. 
 
 We may notice that the QRPA values of $\chi_F$ are always
 smaller with the quenched value $g_A = 1.0$ compared to the unquenched value $g_A = 1.27$.
 That trend continues when the amount of quenching is increased, e.g. to $g_A =0.8$ where 
 $\chi_F$ values are really quite close to -1/3. However, the question of quenching of the $0\nu\beta\beta$
 matrix elements remains  open, and in particular how to treat it properly in the QRPA goes beyond
 the scope of the present paper. \\
 

\begin{table*}[htb] 
\caption{ Ratio $\chi_F = M^{0\nu}_F/M^{0\nu}_{GT}$ (see our definition
of $\chi_F$ in the Eq. (\ref{eq:mgt}) ) in ISM \cite{ISM}, 
QRPA-A, QRPA-B (present work, $g_A=1.00$ and $g_A=1.27$ side by side), QRPA-A results are with
the Argonne V18 potential, QRPA-B with the Bonn CD potential),  IBM-2 \cite{barea13}, and QRPA-JyLa \cite{qrpaj}.
}
\label{tab:chif}    
\centering 
\renewcommand{\arraystretch}{1.1}  
\begin{tabular}{lcccccc}
\hline \hline 
Nucleus & & \multicolumn{5}{c}{$\chi_F$} \\ \hline
          & &  ISM     &    QRPA-A    &      QRPA-B      &  IBM &   QRPA-JyLa  \\ \hline \hline
$^{48}Ca$ & &  -0.22   & -0.42, -0.51  &	-0.42, -0.51    &  -0.68 &    -0.90$^a$   \\
$^{76}Ge$ & &  -0.16   & -0.31, -0.34  &	-0.30, -0.34    & -0.61 &	-0.35	\\
$^{82}Se$ & &  -0.16   & -0.33, -0.36  &	-0.33, -0.36    & -0.68 &	-0.45	\\
$^{96}Zr$ & &     ---   & -0.45, -0.52  &	-0.45, -0.51    & -0.10 &	-0.69	\\
$^{100}Mo$ & &     ---     & -0.44, -0.49  &	-0.43, -0.48    & -0.10 &	-0.64	\\
$^{110}Pd$ & &  -0.25   & -0.39, -0.44  &	-0.38, -0.46    & -0.05 &	-0.61	\\
$^{116}Cd$ & &  -0.30   & -0.43, -0.48  &	-0.42, -0.46    & -0.10 &	-0.45	\\
$^{124}Sn$ & &  -0.20   &    ~~--- ~ -0.43  & ~~--- ~~  -0.41  & -0.56 & 	-0.68	\\
$^{128}Te$ & &  -0.20   & -0.39, -0.43  &	-0.38, -0.41    & -0.55 & 	-0.60	\\
$^{130}Te$ & &  -0.20   & -0.41, -0.44  &	-0.39, -0.42     & -0.55 & 	-0.60	\\
$^{136}Xe$ & &  -0.20   & -0.38, -0.41  &	-0.36, -0.39     & -0.55 &	-0.60	\\
\hline \hline  
\end{tabular}  \\
$^a$ Ref.\cite{Suh93}
\end{table*}

\section{Conclusions}

By separating the particle-particle neutron-proton interaction into its isovector and
isoscalar parts, and renormalizing them each separately with its own fitted parameters 
$g_{pp}^{T=1}$ and $g_{pp}^{T=0}$, we have achieved the partial restoration 
of the isospin symmetry and fulfillment of the requirement that $M^{2\nu}_F = 0.0$.
This has been done essentially without introducing new parameters, since   $g_{pp}^{T=1} \approx g_{pair}$
as required by the isospin symmetry of the particle-particle force. At the same time the
isoscalar parameter $g_{pp}^{T=0}$ is fitted from the requirement that the 
calculated $2\nu\beta\beta$ half-life is the same as its experimental value. The
resulting  $g_{pp}^{T=0}$ is then almost  the same one as with the old
parametrization with the single $g_{pp}$ value.

When the new parametrization of the particle-particle renormalization constants
is used in the QRPA evaluation of the $0\nu\beta\beta$ nuclear matrix elements,
a substantial reduction of the Fermi part, $M^{0\nu}_F$, is observed, while the
Gamow-Teller and Tensor parts remain essentially unaffected. The full
matrix elements $M^{0\nu}$ are reduced by $\sim$ 10 -20\% as seen in Fig. \ref{fig:me0n}. We believe that
such reduction, which also brings the ratio $\chi_F$ closer to $\approx -1/3$, nearer to its
value in the isospin conserving nuclear shell model values, is realistic, and should
be used in the future application of the QRPA and its generalizations.

\section*{Acknowledgments}

 Useful discussions with Kazuo Muto are appreciated. The work of P.V.\ was
partially supported by the US DOE Grant DE-FG02-92ER40701. F. \v{S}. 
acknowledges the support by the VEGA Grant agency
of the Slovak Republic under the contract No. 1/0876/12 and by the Ministry 
of Education, Youth and Sports of the Czech Republic under contract LM2011027.
F.\v{S}. and A. F. thank the Deutsche Forschungsgemeinschaft for support under
contract 436SLK113/14/0-1.



\begin{thebibliography}{99}

\bibitem{AEE} F. T. Avignone, S. R. Elliott and J. Engel, Rev. Mod. Phys. {\bf 8
0}, 481 (2008).

\bibitem{us1} V. A. Rodin, A. Faessler, F. \v{S}imkovic and P. Vogel,
Phys. Rev. C{\bf 68}, 044302(2003).
\bibitem{us2} V. A. Rodin, A. Faessler, F. \v{S}imkovic and P. Vogel,
Nucl. Phys. {\bf A766}, 107 (2006), and erratum {\bf A793}, 213 (2007).
\bibitem{us3} 
F. \v{S}imkovic, A. Faessler, V. A. Rodin, P. Vogel, and J. Engel, Phys. Rev. C{\
bf 77}, 045503(2008).

\bibitem{us5} F. \v{S}imkovic, R. Hod\'{a}k, A. Faessler, and P. Vogel,
Phys. Rev. C{\bf 83}, 015502(2011).

\bibitem{GCM} T. R. Rodriguez and G. Martinez-Pinedo, Phys. Rev. Lett. {\bf 105}, 
252503(2010).

\bibitem{PHFB} K. Chatuverdi, R. Chandra, P. K. Rath, P. K. Raina and J. G. Hirsch, Phys. Rev. C{\bf 78}, 054302 (2008).

\bibitem{IBM2} J. Barea and F. Iachello, Phys. Rev. C{\bf 79}, 044301 (2009).

\bibitem{Vadim11} V. Rodin and A. Faessler, Phys. Rev. C{\bf 84}, 014322 (2011).

\bibitem{ISM} E. Caurier, F. Nowacki, and A. Poves, Int. J. Mod. Phys. {\bf E16}, 552 (2007);
 J. Men\'endez, A. Poves, E. Caurier, and F. Nowacki, Nucl. Phys. A {\bf 818}, 139 (2009);
 E. Caurier, F. Nowacki and A. Poves; Eur. Phys. J.{\bf A36}, 195 (2008).
 
 \bibitem{Suh88} J. Suhonen, T. Taigel and A. Faessler, Nucl. Phys. {\bf A486}, 91 (1988).
 
\bibitem{VZ86} P. Vogel and M. R. Zirnbauer, Phys. Rev. Lett. {\bf 57}, 3148 (1986);
J. Engel, P. Vogel and M. R. Zirnbauer, Phys. Rev. C{\bf 37}, 731 (1988).

\bibitem{Sim09} F. \v{S}imkovic, A. Faessler, H. M\"{u}ther, V. Rodin,
and M. Stauf,  Phys. Rev. C{\bf 79}, 055501 (2009).

\bibitem{Barab} A. S. Barabash,  Phys. Rev. C{\bf 81}, 035501 (2010).

\bibitem{EXO} N. Ackerman {\it et al.} , Phys. Rev. Lett. {\bf 107}, 212501 (2011).

\bibitem{ZEN} A. Gando {\it et al.} Phys. Rev.C{\bf 85}, 045504 (2012).

\bibitem{Pd110} D. Fink {\it et al.}, Phys. Rev. Lett. {\bf 108}, 062502 (2012).

\bibitem{Cau08} E. Caurier, J. Menendez, F. Nowacki and A. Poves, Phys. Rev. Lett. {\bf 100}, 052503 (2008).

\bibitem{Horoi} M. Horoi, arXiv: 1210.6680.

\bibitem{barea13} J. Barea, J. Kotila, and F. Iachello, Phys. Rev. C{\bf87}, 014315 (2013); 
arXiv:1301.4203 [nucl-th]. 

\bibitem{qrpaj} J. Suhonen, private communications as quoted in \cite{barea13}.

\bibitem{Suh93} J. Suhonen, J. Phys. G{\bf 19}, 139 (1993).

\end{thebibliography}
\end{document}